\documentclass[twocolumn,nodate,showpacs,preprintnumbers,nofootinbib,amsmath,amssymb,aps,prd]{revtex4}

\usepackage{amssymb}
\usepackage{amsmath,amssymb,amsfonts,amsthm}
\usepackage{graphicx}

\def\be{\nopagebreak[3]\begin{equation}}
\def\ee{\end{equation}}
\def\ba{\nopagebreak[3]\begin{eqnarray}}
\def\ea{\end{eqnarray}}

\def\lp{\ell_{\rm Pl}}

\def\f{\frac}
\def\mpl{m_{\rm Pl}}

\def\rcr{\rho_{\rm max}}
\def\t{\tilde}
\def\h{\hat}
\def\sint{\textstyle{\int}}
\def\tr{\rm Trun}
\def\pphi{p_{(\phi)}}
\def\x{\vec{x}}
\def\vk{\vec{k}}
\def\p{\mathfrak{p}}

\def\T{\mathcal{T}}
\def\Q{\mathcal{Q}}

\def\H{\mathcal{H}}
\def\B{{\rm B}}

\begin{document}

\title{A Quantum Gravity Extension of the Inflationary Scenario}

\author{Ivan Agullo}
\affiliation{Institute for
Gravitation and the Cosmos \& Physics
  Department, Penn State, University Park, PA 16802, U.S.A.}
\author{Abhay Ashtekar}
\affiliation{Institute for Gravitation and the Cosmos \& Physics
  Department, Penn State, University Park, PA 16802, U.S.A.}
\author{William Nelson}
\affiliation{Institute for Gravitation and the Cosmos \& Physics
  Department, Penn State, University Park, PA 16802, U.S.A.}

\begin{abstract}

Since the standard inflationary paradigm is based on quantum field
theory on classical space-times, it excludes the Planck era. Using
techniques from loop quantum gravity, the paradigm is extended to a
self-consistent theory from the Planck scale to the onset of slow
roll inflation, covering some 11 orders of magnitude in energy
density and curvature. This pre-inflationary dynamics also opens a
small window for novel effects, e.g. a source for non-Gaussianities,
which could extend the reach of cosmological observations to the
deep Planck regime of the early universe.

\end{abstract}


\maketitle

The inflationary paradigm has had remarkable success in accounting
for the inhomogeneities in the cosmic microwave background (CMB)
that serve as seeds for the large scale structure of the universe.
However it has certain conceptual limitations from particle physics
as well as quantum gravity perspectives. For example: i) The
physical origin of the inflaton and its properties remains unclear;
ii) Since the background geometry and matter satisfy Einstein's
equations, the big bang singularity persists \cite{bgv}; iii) One
ignores pre-inflationary dynamics and simply requires that
perturbations be in the Bunch Davies (BD) vacuum at the onset of the
slow roll; and, iv) When evolved back in time these perturbative
modes acquire trans-Planckian frequencies and the underlying
framework of quantum field theory on classical space-times becomes
unreliable. Here we will not address any of the particle physics
issues. Rather, we focus on the incompleteness related to quantum
gravity and show that this limitation can be overcome. In addition,
we find that pre-inflationary dynamics can produce certain
deviations from the BD vacuum at the onset of inflation, leading  to
novel effects which could be seen, e.g., in non-Gaussianities
through future measurements of the halo bias and the `$\mu$-type
distortions' in the CMB \cite{halo-bias}.

Loop quantum gravity (LQG) offers a natural framework to address
these issues because effects of its underlying quantum geometry
dominate at the Planck scale, leading to singularity resolution in a
variety of cosmological models, including some that admit
anisotropies and inhomogeneities \cite{asrev}. Even though LQG is
still incomplete, notable advances have occurred
---e.g., in cosmology, analysis of black holes, and a derivation of
the graviton propagator--- by using the following strategy:
\emph{First carry out a truncation of the classical theory geared to
the given physical problem and then use LQG techniques to construct
the quantum theory} \cite{zakopane}. For inflation, then, we are led
to focus just on first order perturbations off the spatially flat
Friedman backgrounds with a scalar field $\phi$. In numerical
simulations we will use the quadratic potential $V = (1/2)
m^2\phi^2$ with $m= 1.21 \times 10^{-6}\mpl$, the value that comes
from the 7 year WMAP data \cite{wmap,as3}. Throughout we use natural
Planck units.

\textbf{The truncated phase space:} We have $\Gamma_{\tr} = \Gamma_o
\times \Gamma_1$ where $\Gamma_o$ is the 4-dimensional phase space
of homogeneous fields, and $\Gamma_1$, of the first order, purely
inhomogeneous perturbations thereon. $\Gamma_o$ is conveniently
coordinatized by the scale factor $a$, the inflaton $\phi$ and their
conjugate momenta. Dynamics on $\Gamma_o$ is generated by the
single, homogeneous, Hamiltonian constraint, $\mathbb{C}_o =0$. On
$\Gamma_1$ the first order constraints can be solved and one can
readily pass to the reduced phase space $\t\Gamma_1$ which we
coordinatize by two tensor modes, collectively denoted by $\T_{\vk}$
in what follows, and the Mukhanov variable $\Q_{\vk}$ representing
the scalar mode (see, e.g., \cite{langlois}). This passage to the
reduced phase space refers only to constraints and does not use any
evolution equations.

Finally, a subtle but conceptually important point is that
dynamics on $\t\Gamma_{\tr} = \Gamma_o \times \t\Gamma_1$, is
\emph{not} generated by a constraint. Rather, the dynamical
flow on $\t\Gamma_{\tr}$ follows the vector field $X^\alpha =
\Omega^{\alpha\beta}_o \partial_\beta \mathbb{C}_o +
\t\Omega^{\alpha\beta}_1\partial_\beta \mathbb{C}_2^\prime$
where $\Omega_o$ and $\t\Omega_1$ are the symplectic structures
on $\Gamma_o$ and $\t\Gamma_1$, and $\mathbb{C}_2^\prime$ is
the part of the second order Hamiltonian constraint in which
only terms that are quadratic in the first order perturbations
are kept. $X^\alpha$ fails to be Hamiltonian on
$\t\Gamma_{\tr}$ because $\mathbb{C}_2^\prime$ depends not only
on perturbations but also on background quantities. However,
given a dynamical trajectory $\gamma_o(t)$ on $\Gamma_o$ and a
perturbation at a point on it, $X^\alpha$ provides a canonical
lift of $\gamma_o(t)$ to the total space $\t\Gamma_{\tr}$,
describing the evolution of that perturbation along
$\gamma_o(t)$.

\textbf{Quantum Kinematics:} Since $\t\Gamma_{\tr} = \Gamma_o \times
\t\Gamma_1$, the total Hilbert space is given by $\H =
\H_o\otimes\H_1$. The Hilbert space $\H_o$ of background fields
consists of wave functions $\Psi_o(a,\phi)$ and its structure is
well understood from loop quantum cosmology (LQC) \cite{asrev}. For
perturbations, we introduce an infrared cutoff so that $\lambda_{\rm
cutoff} \ge \lambda_o$, the size of the observable universe.
Physically, this amounts to `absorbing modes with $\lambda
> \lambda_{\rm cutoff}$ in the background'. Then there is a natural
Hilbert space $\H_1$ on which perturbations $\h{\Q}_{\vk}$ and
$\h{\T}_{\vk}$ act. It admits an infinite dimensional sub-space
of 4th order adiabatic states \cite{pf} which are invariant
under spatial translations, often called \emph{`vacua'}. $\H_1$
is generated by excitations on any one of them. (For an
alternate characterization see \cite{iberian}.) Note however
that, in contrast to quantum field theory on strictly
stationary space-times, $\H_1$ does not have a preferred vacuum
state, or a canonical notion of particles.

The key difference from standard inflation is that quantum fields
$\h{\Q}_k, \h{\T}_k$ now propagate on a \emph{quantum geometry}
represented by $\Psi_o(a,\phi)$ rather than on a classical Friedmann
solution $(a(t), \phi(t)$). These quantum geometries are all
regular, free of singularities. Thus, by construction, the framework
encompasses the Planck regime.

Now, the quantum geometry underlying LQG is subtle
\cite{zakopane,ttbook}. For example, while there is a minimum
non-zero eigenvalue of the area operator, there is no such
minimum for the volume operator, although its eigenvalues are
also discrete. In the present truncated theory, perturbative
modes with arbitrarily high frequencies are allowed even though
there is a quantum geometry $\Psi_o$ in the background. By
itself, this is not a problem. In our homogeneous sector, for
example, the inflaton momentum $\pphi$ can be arbitrarily large
but still the energy density $\rho$ is bounded above by $\rcr
\sim\, 0.41 \rho_{\rm Pl}$ \cite{asrev}. The \emph{real}
trans-Planckian issue for us is whether the \emph{energy
density} in perturbations remains (not only bounded but) small
compared to the background all the way back to the bounce. Only
then would we be assured of a self consistent solution,
justifying our truncation which ignores the back reaction.
Otherwise one would have to await a full quantum gravity
theory.

\textbf{Quantum dynamics:} Since the classical dynamics on
$\t\Gamma_{\tr}$ is not generated by a constraint, contrary to
what is often done, one cannot recover quantum dynamics for the
\emph{total system} by imposing a quantum constraint. As in the
classical theory, we can do this only in the homogeneous sector
$\H_o$ and we then have to `lift' the resulting quantum
trajectory to the full $\H$. On $\H_o$ one can follow the
standard procedure in LQC. It again leads us to reinterpret the
quantum Hamiltonian constraint $\h{\mathbb{C}}_o \,\Psi_o =0$
as an `evolution' equation, $-i\hbar\partial_\phi
\Psi_o(a,\phi) = \h{H}_o\Psi_o (a, \phi)$, with respect to the
relational or emergent time variable $\phi$ generated by a time
dependent Hamiltonian $\h{H}_o$ \cite{aps4,asrev}. In this
analysis one encounters certain technical complications because
of the presence of the potential $V(\phi)$. Their origin and
resolution is analogous to that in the case where $V(\phi)=0$
but there is a positive cosmological constant \cite{pa}.

In the $V(\phi) =0$ case, detailed investigations have shown that
wave functions $\Psi_o(a,\phi)$ of physical interest remain sharply
peaked even in the Planck era and follow quantum corrected effective
trajectories. For $V(\phi)= (1/2)m^2\phi^2$, solutions to effective
equations continue to undergo a bounce when $\rho=\rcr$ and to agree
with general relativity for $\rho \lesssim 10^{-3}\rcr$. However,
because of computational limitations, so far the quantum wave
functions $\Psi_o(a,\phi)$ have been calculated only when the bounce
is kinetic energy dominated \cite{aps4}. In this case, the peaks of
wave functions of interest again follow the effective trajectories
as expected. We restrict ourselves to background quantum geometries
$\Psi_o(a,\phi)$ with this property. Each of them provides a
probability amplitude for various classical space-time geometries to
occur. They are peaked not on classical Friedmann solutions but
rather on quantum corrected bouncing solutions. Furthermore, there
are fluctuations around these peaks. \emph{The challenge is to
capture the effects of this background quantum geometry
$\Psi_o(a,\phi)$ on the dynamics of perturbations.}

To meet it, we use the conceptual framework of quantum field theory
on cosmological \emph{quantum} geometries, introduced in \cite{akl}.
An extension of that framework to incorporate an infinite number of
modes, with appropriate regularization and renormalization, provides
the dynamical equation for states $\psi(\Q_{\vk}, \T_{\vk})$ of
perturbations on the background quantum geometry $\Psi_o$. A key
result is that this evolution \emph{is equivalent to that of test
perturbations propagating on a dressed, effective, smooth metric}
\vskip0.1cm
\centerline{$\tilde{g}_{ab} dx^a dx^b\, \equiv\, d\tilde{s}^2 \, =
\, \tilde{a}^2(\phi)\, (-d \tilde{\eta}^2 + d\x^2)$}
\vskip0.1cm
\noindent where the dressed scale factor $\t{a}$ and the dressed
conformal time $\t\eta$ are given by
\be \tilde{a}^4 = {\langle \hat{H}_o^{-\f{1}{2}}\, \hat{a}^4(\phi)\,
\hat{H}_o^{-\f{1}{2}}\rangle}{\langle \hat{H}_o^{-1}\rangle}^{-1};
\,\, d\tilde{\eta} = \tilde{a}^2(\phi)\, \langle
\hat{H}_o^{-1}\rangle\, d\phi .\nonumber \ee
This result is exact within our truncation scheme. It shows that the
propagation of perturbations is sensitive to properties of the state
$\Psi_o$ even beyond the quantum corrected effective geometry
followed by its peak; it also senses quantum fluctuations around
this peak. However, interestingly, this dependence is neatly coded
in just two `dressed' quantities, $\t\eta$ and $\t{a}$. This is
analogous to the fact that although light propagating in a medium
interacts with its atoms, the net effect can be captured in just a
few parameters such as the refractive index.

This result greatly simplifies our task conceptually and enables us
to use the technical tools of mode by mode regularization and
renormalization from the well developed adiabatic scheme of quantum
field theory on classical cosmological space-times \cite{pf}. For
tensor modes, for example, one obtains the following evolution
equation \vskip-0.5cm
\ba &&i\hbar\partial_{\t\eta} \psi(\T_{\vk}, \t\eta) = \hat{H}_1
\psi(\T_{\vk}, \t\eta) \nonumber\\
&&\equiv \f{1}{2}\, \sint d^3{k}\,\, \Big[\f{4\kappa}{ \t{a}^{2}}\,
|\hat{\p}_{\vk}|^2 +\f{k^2\,\t{a}^2}{4\kappa}\, |\h{\T}_{\vk}|^2\,
\,\, - C_k(\t\eta)\Big]\, \psi(\T_{\vk},\t\eta) \label{qevo2}
\nonumber\ea
where $\h\p_{\vk}$ is the momentum conjugate to $\h{\T}_{\vk}$,\,
$\kappa=8\pi G$ and $C_k(\t\eta)$ are c-numbers,  derived from the
4th order adiabatic regularization that depend only on $k=|\vk|$.

\textbf{Initial conditions:} Since the big bang is replaced by
the big bounce, it is natural to specify initial conditions at
the bounce. The initial state can be taken to be of the form
$\Psi_o \otimes \psi$ because perturbations are treated as test
fields. This tensor product form prevails so long as the back
reaction remains negligible during evolution. To specify the
initial condition for $\Psi_o$ let us first recall that, in
effective LQC, all dynamical trajectories enter a slow roll
phase compatible with the 7 year WMAP data unless $\phi_{\B}$,
the value of the inflaton at the bounce, lies in a \emph{very
small} region $R$ of the constraint surface \cite{as3}. We will
assume that, at the bounce, the background quantum state
$\Psi_o$ is sharply peaked at a point on the constraint surface
anywhere outside this $R$. In this sense the initial data for
$\Psi_o$ is generic. For perturbations, we assume that the
initial $\psi$ is a 4th order adiabatic `vacuum' such that the
expectation value of the renormalized energy density in $\psi$
is negligible compared to that in the background. This is a
large class of initial data for test fields, selected by
general symmetry requirements.

Physically, we are assuming `initial quantum homogeneity' i.e.,
requiring that the region which expands to become the observable
universe is homogeneous at the bounce except for `vacuum
fluctuations'. While this is a strong restriction, it may be
naturally realized in LQG because: i) In solutions of interest, the
observable universe has a radius  $\lesssim 10 \lp$ at the bounce;
and, ii) The strong repulsive force due to quantum geometry that
causes the bounce has a `diluting effect'. It could make this
`quantum homogeneity' generic, `washing out' the memory of the
pre-bounce dynamics at the scale $\lesssim 10\lp$.

Our remaining task is two fold: i) starting from these initial
conditions, calculate the power spectrum for scalar and tensor modes
at the end of the slow roll inflation; and, ii) verify if the back
reaction continues to remain negligible all the way to the onset of
the slow roll so that our initial truncation is a self consistent
approximation.
\begin{figure}[]
 \begin{center}
  \includegraphics[scale=0.65]{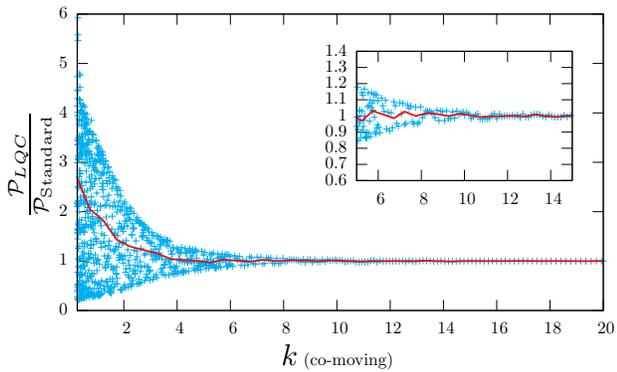}
  \caption{\label{fig1} Ratio of our LQG power spectrum for scalar
  perturbations to the standard inflationary power spectrum. The (blue)
  crosses denote the data points.
  For small $k$, the ratio oscillates rapidly with $k$. The solid
  (red) curve shows averages over bins of width $\Delta k\!=\!0.5\,
  {\lp}^{-1}$. The inset shows a blow-up of the interesting region
  around $k=9$.}
 \end{center}
\end{figure}

\textbf{Power spectrum:} As noted above, in bounces with
kinetic energy domination on which we focus, the quantum state
$\Psi_o(a,\phi)$ is known to remain sharply peaked on effective
trajectories. Therefore, in numerical simulations a `mean
field' approximation was made by replacing $\tilde{a}(\phi),
\tilde{\eta}(\phi)$ by the mean values of these operators. For
the background,  several simulations were carried out with
$\phi_{\B}$ in $(0.93\mpl,\, 1.5\mpl)$, which, as we will see
below, is the most interesting range. For perturbations, we
used three different initial states $\psi$ in the class
specified above. The power spectrum at the end of inflation was
computed in each case for both scalar and tensor modes. Results
are all very similar. FIG.1 shows how the LQC scalar power
spectrum relates to the prediction of standard inflation for
the case where $\phi_B = 1.15 \mpl$, and the initial state
$\psi$ is the `obvious' or `standard' 4th order adiabatic
vacuum. We found that the plot is largely insensitive to
choices of initial conditions within the class used in our
simulations.

Recall, however, that the 7 year WMAP data \cite{wmap} covers only a
window $(k_{\rm min} \approx k^\star/8.58, \,\, k_{\rm max} \approx
2000 k_{\rm min})$ in the co-moving $k$ space. 
Here the reference mode $k^\star$ is the one that exits the Hubble
radius at time $\t\eta_{k^\star}$ when the Hubble parameter is given
by $H(\t\eta_{k^\star}) = 7.83 \times 10^{-6} \mpl$. In FIG.1,
numerical values of the \emph{co-moving} $k$ were calculated using
the scale factor convention $a_{\B}\!=\! 1$, rather than $a_{\rm
today}\! = \!1$. (The \emph{physical} wave numbers are of course
convention independent). In each simulation, we first locate the
scale factor $\t{a}(\t\eta_{k^\star})$ by setting $H =
H(\t\eta_{k^\star})$, and then determine $k^\star$ via $k^\star =
\t{a}(\t\eta_{k^\star}) H(\t\eta_{k^\star})$. Since we have
$\t{a}_{\B}\! =\!1$, values of $\t{a}(\t\eta_{k^\star})$ and
$k^\star$ depend on the pre-inflationary background dynamics which
turns out to be governed entirely by $\phi_{\B}$. Therefore, in
FIG.1 \emph{the observationally relevant window depends on the value
of $\phi_{\B}$}, moving steadily to right as $\phi_{\B}$ increases.

The plot has two interesting features. First, \emph{the LQG power
spectrum is virtually indistinguishable from that of standard
inflation} if $k_{\rm min} \gtrsim 9 \mpl$. This occurs when
$\phi_\B \gtrsim 1.2 \mpl$. Second, for smaller values of $k_{\rm
min}$, the observational window admits modes for which the two power
spectra are noticeably different. For concreteness, let us set
$\phi_{\B} = 1.15\mpl$. Then $k_{\rm min} \simeq 1.07\mpl$ and these
modes correspond to $\ell \lesssim 30$ in the WMAP angular
decomposition for which observational error bars are large.
Therefore the LQG power spectrum is also viable but the
\emph{predicted quantum state of perturbations at the onset of
inflation is not the BD vacuum} for $\phi_{\B} < 1.2\mpl$.

\textbf{Self consistency:} Whether the test field approximation
continues to hold in the Planck regime is an intricate issue and had
not been explored before. FIG.2 shows that we have obtained explicit
self consistent solutions $\psi$ in which the renormalized energy
density in perturbations remains low compared to the background
\emph{all the way from the bounce to the onset of inflation.} (Here,
we have set $k_{\rm cutoff}\! =\! k_o\! =\! 30 \mpl$, which
corresponds to $\phi_B \approx 1.23\mpl$.) Furthermore, there is an
analytical argument showing that every such $\psi$ admits a
well-defined neighborhood (in the infinite dimensional space of the
4th order adiabatic `vacua') with the same property. Thus, for
$\phi_\B \gtrsim 1.23\mpl$, our truncated framework admits a rich
set of self consistent solutions. Furthermore, in each of them
$\psi$ has extremely small excitations over the BD vacuum at the
onset of slow roll. \emph{These solutions provide viable extensions
of the standard inflationary scenario all the way to the Planck
scale.}

What about the small but interesting window $\phi_{\B} < 1.2 \mpl$?
So far we only have upper bounds for the renormalized energy density
in perturbations and these are far from being optimal. Therefore we
do not yet have an explicit solution establishing the validity of
the test field approximation in this window.
\begin{figure}
 \begin{center}
  \includegraphics[scale=0.65]{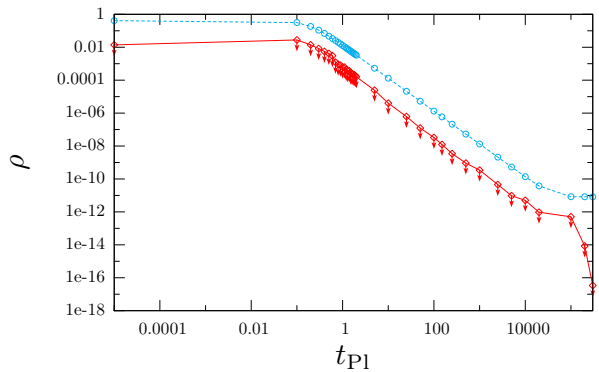}
  \caption{\label{fig2} Energy
  density in the background (upper (blue) curve) and an upper bound on
  the \emph{renormalized} energy density in perturbations (lower (red) curve)
  are plotted against time from the bounce to the onset of slow
  roll, using Planck units. The test field approximation holds across
  a change of over 11 orders of magnitude in both quantities.}
 \end{center}
\end{figure}

\textbf{Summary and discussion:} Using LQG ideas and techniques, we
have extended the inflationary paradigm all the way to the deep
Planck regime. At the big bounce, one can specify natural initial
conditions for the quantum state $\Psi_o$ that encodes the
background homogeneous quantum geometry, as well as for $\psi$ that
describes the quantum state of perturbations. There is a precise
sense in which generic initial conditions for the background lead to
a slow roll phase compatible with the 7 year WMAP data \cite{as3}.
We have shown that there is a large set of initial data for $\psi$
such that:\, i) at the onset of slow roll, $\psi$ is
\emph{extremely} close to the BD vacuum, and, ii) the test field
approximation behind the truncation strategy is self consistent.
Each of these solutions provides a viable quantum gravity completion
of the standard inflationary paradigm. However particle physics
issues still remain.

In addition, there exists a narrow window, $\phi_{\B} < 1.2
\mpl$ for which the quantum state $\psi$ at the onset of
inflation has an appreciable number of `BD particles' (but
within the current observational limits). The physical origin
of this effect can be explained in terms of the new scale $k_M$
defined by the \emph{universal} value of the scalar curvature
at the bounce. Excitations with $k\lesssim k_M$ are created in
the Planck regime near the bounce. It turns out that if the
number $N$ of e-foldings in $\t{a}$ between the bounce and
$\t\eta = \t\eta_{k^\star}$ is less than 15, then $k_{\rm min}<
k_M$, whence some of these modes would be in the window
accessible through CMB. $N <15$ corresponds to $\phi_B \lesssim
1.2 \mpl$, precisely the regime in which the LQC power spectrum
is different from the BD vacuum. Future measurements should be
sensitive to such deviations \cite{halo-bias}. If they are
observed at the scale $k=k_M$, the parameter space of initial
conditions for $\Psi_o$ would be tightly squeezed, making much
more detailed predictions feasible. In this sense, the
framework expands the reach of observational cosmology all the
way to the deep Planck regime. This general argument also shows
that the pre-inflationary dynamics has negligible effect for
modes with $k \gg k_M$ because their physical wave lengths turn
out to be smaller than the curvature scale throughout the
evolution. This explains the very close agreement between the
LQC and the standard power spectrum at high $k$.

Finally, interesting and complementary investigations of LQG
dynamics between the bounce and the onset of slow roll have appeared
in the literature recently (see, especially, \cite{lqc}). The
distinguishing features of our analysis are: i) It is based on the
general truncation strategy that has proven to be successful in
other problems; ii) It provides a systematic approach to quantum
dynamics, made necessary by the fact that the classical evolution is
not generated by a constraint on  $\Gamma_{\tr}$; iii) The treatment
of initial states has been stream-lined; and, most importantly, iv)
While issues of regularization of the Hamiltonian operator
$\h{H}_1$, adiabatic renormalization of energy density, and
consistency of the test field approximation were ignored so far,
they have now been addressed using quantum field theory on quantum
geometries. Details and subtleties which could not be included here
will be discussed in two forthcoming articles.

\emph{Acknowledgments:} This work is supported in part by the NSF
grant PHY-1205388.

\end{document}